\documentclass[runningheads]{llncs}
\usepackage[T1]{fontenc}

\usepackage{cite}
\usepackage{amsmath,amssymb,amsfonts}
\usepackage{algorithmic}
\usepackage{graphicx}
\usepackage{textcomp}
\usepackage[table,xcdraw]{xcolor}
\usepackage{multirow}
\usepackage{tabularx}
\usepackage{hyperref}
\usepackage{array}
\newcolumntype{P}[1]{>{\centering\arraybackslash}p{#1}}

\usepackage{tikz}
\usetikzlibrary{arrows.meta, positioning, calc}

\usepackage{booktabs}
\newcommand{\graycell}{\cellcolor{gray!10}}

\usepackage{rotating}
\usepackage{adjustbox}
\usepackage{amssymb}

\usepackage{orcidlink}
\usepackage{marvosym}
\newcommand{\emailicon}{\textsuperscript{(\Letter)}}

\usepackage{enumitem}
\newlist{enumerate*}{enumerate}{1}
\setlist[enumerate*]{label=\textit{(\roman*)}, itemjoin={{, }}, itemjoin*={{, and }}}

\begin{document}

\title{Generative AI in Simulation-Based \\ Test Environments for Large-Scale Cyber-Physical Systems: An Industrial Study}

\titlerunning{Generative AI in Simulation-Based Test Environments for Large-Scale CPSs}

\author{
Masoud Sadrnezhaad\inst{1}\emailicon\orcidlink{0000-0002-3437-1681} \and
José Antonio Hernández López\inst{2}\orcidlink{0000-0003-2439-2136} \and
Torvald Mårtensson\inst{1}\orcidlink{0000-0003-1438-0182} \and
Dániel Varró\inst{1}\orcidlink{0000-0002-8790-252X}
}

\authorrunning{Sadrnezhaad et al.}

\institute{
Linköping University, Linköping, Sweden\\
\email{\{masoud.sadrnezhaad, torvald.martensson, daniel.varro\}@liu.se}
\and
University of Murcia, Murcia, Spain\\
\email{joseantonio.hernandez6@um.es}
}

\maketitle

\begin{abstract}
Quality assurance for large-scale cyber-physical systems relies on sophisticated test activities using complex test environments investigated with the help of numerous types of simulators. As these systems grow, extensive resources are required to develop and maintain simulation models of hardware and software components, as well as physical environments. Meanwhile, recent advances in generative AI have led to tools that can produce executable test cases for software systems, offering potential benefits such as reducing manual efforts or increasing test coverage. However, the application of generative AI techniques to simulation-based testing of large-scale cyber-physical systems remains underexplored. To better understand this gap, this study captures practitioners’ perspectives on leveraging generative AI, based on a cross-company workshop with six organizations. Our contribution is twofold: (1) detailed, experience-based insights into challenges faced by engineers, and (2) a research agenda comprising three high-priority directions: (a) AI-generated scenarios and environment models, (b) simulators and AI in CI/CD pipelines, and (c) trustworthiness in generative AI for simulation. While participants acknowledged substantial potential, they also highlighted unresolved challenges. By detailing these issues, the paper aims to guide future academia-industry collaboration towards the responsible adoption of generative AI in simulation-based testing.

\end{abstract}

\section{Introduction}
\noindent\textbf{Background}
Quality assurance for large-scale cyber-physical systems (CPSs), such as aircraft or road vehicles, requires various test environments at different levels of granularity and integration \cite{DBLP:journals/access/ZhouGHY18}. Based on detailed models of hardware components (e.g., a model of fuel injector in a car), software components (e.g., a model of fuel injection controller) and the physical environment (e.g., a model of fuel viscosity), such test environments investigate the system's behavior using a multitude of simulation tools \cite{DBLP:journals/access/ZhouGHY18}.

In previous work, we have repeatedly touched upon problems and challenges related to test environments. In a study of continuous integration impediments, “reliability of test environments” was identified as one of twelve factors that could enable more frequent integration of software \cite{maartensson2017continuous}. In another study, industry practitioners described how better models of physical systems (e.g., a simulator model of the fuel system in a car) were increasingly important, as testing with real hardware is too expensive \cite{https://doi.org/10.1002/stvr.1839}. Some interviewees described constructing good models of physical systems as an additional challenge (e.g., simulating how a liquid flows through the pipes). As systems become larger and more complex, the implementation and maintenance costs of simulator models increase over time, necessitating new solutions to reduce costs and save time.

\noindent\textbf{Research question}
Thanks to recent advances in generative AI techniques, frameworks based on large language models, such as ChatGPT or Copilot, are capable of writing software that solves complex problems~\cite{zhuo2024bigcodebench, DBLP:conf/vecos/PatilUN24}, generating test cases~\cite{alshahwan2024automated}, and evaluating test results~\cite{DBLP:conf/profes/KarlssonLSS24}. 
Generative AI techniques can be applied to simulations in test environments for large-scale cyber-physical systems, potentially reducing costs and saving time. In the companies the authors have worked with as researchers, practitioners have begun exploring AI simulation techniques. However, these efforts often lack a structured and holistic approach. To address this gap, the paper aims to answer the following question: \textit{What is the potential of generative AI techniques in simulation-based test environments for large-scale cyber-physical systems according to industry practitioners?}

\noindent\textbf{Contribution}
We conducted an in-depth workshop with six companies working on large-scale CPSs, followed by a thematic coding analysis of all the material. Based on this evidence, our contribution consists of two main parts. (1) We provide \emph{industry-grounded insights} into the challenges practitioners face in simulation-based testing. Concrete obstacles include large-scale data handling and scenario evaluation and selection, which offer an up-to-date, experience-based understanding of barriers that limit scalable simulation-based testing. (2) 
We propose an \emph{actionable research agenda} that outlines three priority directions for the use of generative AI in this context, namely, (a) AI-generated scenarios and environment models, (b) integration of simulators and AI into CI/CD pipelines, and (c) ensuring trustworthiness in generative AI for simulation. 

\noindent\textbf{Relevance and novelty}
In this paper, we provide researchers with clear, high-impact directions for future work and support practitioners in making informed decisions about adoption, thereby guiding the responsible and effective use of generative AI in large-scale simulation-based testing.

To the best of our knowledge, this is the first paper that investigates the \emph{industrial practitioners' viewpoint} on the use of \emph{generative AI} for \emph{simulation-based test environments} in \emph{large-scale cyber-physical systems}. Existing studies on using AI in a CPS context \cite{DBLP:journals/spe/CederbladhEMS24,DBLP:journals/tcps/LeeCJKHUKP25,DBLP:journals/corr/abs-2406-17112} do not specifically target simulation environments, while the authors of \cite{DBLP:journals/corr/abs-2501-04410} assess simulation and generative AI, but not in a CPS context. 

\noindent\textbf{Organization} \autoref{sec:method} explains the research method. \autoref{sec:literature} reviews recent literature and establishes the background for generative AI in simulation-based testing of CPSs. \autoref{sec:summaries} summarizes the workshop, detailing the individual presentations and group discussions of the participating companies. \autoref{sec:thematic} provides a thematic coding analysis of companies' input. Threats to validity are discussed in \autoref{sec:threats}, and the paper is concluded in \autoref{sec:conclusion}.

\section{Research method} \label{sec:method}

\begin{figure}[b]
\centering
\begin{tikzpicture}[
    node distance=0.5cm and 0.5cm,
    every node/.style={rectangle, draw, align=center, minimum width=3cm, minimum height=1cm, font=\scriptsize},
    outcome/.style={minimum height=0.2cm, font=\scriptsize},
    >=Stealth
  ]
  \node (litRev) {\textbf{Phase 1:}\\Preliminary study};
  \node (workshop) [right=of litRev] {\textbf{Phase 2:}\\Exploration of\\practitioners' perspectives};
  \node (thematic) [right=of workshop] {\textbf{Phase 3:}\\Thematic coding analysis};

  \draw[->, thick] (litRev) -- (workshop);
  \draw[->, thick] (workshop) -- (thematic);

  \node[outcome, minimum width=1.8cm] (s1out) [below=of litRev] {Taxonomy and\\ insights};
  \node[outcome, minimum width=1.8cm] (s2out) [below=of workshop] {Company presentations \\ and group discussions};
  \node[outcome, minimum width=1.8cm] (s3out) [below=of thematic] {Codes and \\ themes};

  \draw[->, thick] (litRev.south) -- (s1out.north);
  \draw[->, thick] (workshop.south) -- (s2out.north);
  \draw[->, thick] (thematic.south) -- (s3out.north);

\end{tikzpicture}
\caption{Research process overview with step labels and outcomes}
\label{fig:research_flowchart}
\end{figure}
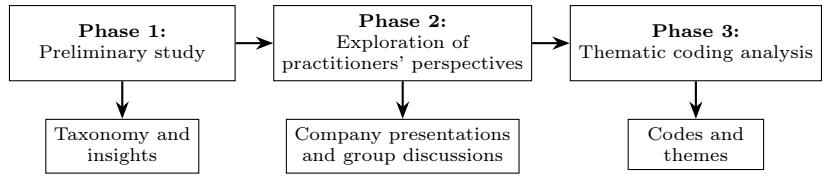

This study has adopted a three-phase methodology shown in \autoref{fig:research_flowchart}. \textbf{Phase~1} involved a preliminary study to ground the research on existing literature and present highlights from the literature to participants as a foundation for discussion in the subsequent workshop. In \textbf{Phase~2}, we explored practitioners' perspectives in a workshop with company presentations and group discussions. These sessions provided insights into current practices, challenges, and expectations. Finally, in \textbf{Phase~3}, we thematically analyzed the workshop data to identify key themes and patterns across practitioners' viewpoints with respect to the role of generative AI in simulation-based test environments for CPS.

\subsection{Phase 1: Preliminary study}
\label{sec:phase-1}

Following Wohlin's guidelines for snowballing in literature studies \cite{wohlin2014guidelines}, we began by establishing the research goal and scope, focusing on the application of generative AI in modeling and simulation workflows for cyber-physical systems. We then constructed a starting set by performing structured keyword searches in databases such as IEEE Xplore. Search terms included combinations of keywords such as “generative AI,” “large language models,” “digital twins,” “simulation,” and “cyber-physical systems.” Candidate papers were selected based on relevance to the research focus, publication quality, and citation impact.

We conducted both backward and forward snowballing from the start set by reviewing reference lists for earlier works and using citation indices to identify newer studies. At each iteration, papers were reviewed and filtered to remove duplicates, non-peer-reviewed sources, and those not aligned with the research focus. This iterative process continued until saturation was reached, that is, no new relevant studies were identified in subsequent snowballing rounds.

The final corpus was classified based on a taxonomy (see \autoref{tab:taxonomy}), and these insights were shared with participants as a foundation for discussion.

\subsection{Phase 2: Exploration of practitioners' perspectives}
\label{sec:phase-2}

To collect industry perspectives, we held a one-day, in-person workshop with 33 participants, including university representatives and software engineers, data scientists, and product managers involved in testing from six companies headquartered in Sweden with global presence. Companies were purposefully selected to represent diverse industry domains where generative AI adoption intersects with the complexities of large-scale systems and domain-specific constraints.

Particularly, Companies 1, 2, and 6 represented the \emph{automotive} sector, Company 3 works in \emph{communications}, Company 4 operates in \emph{aerospace}, and Company 5 specializes in \emph{cybersecurity}. The companies range in size from small (around 60 employees) to large enterprises with up to 95,000 employees.

After an initial presentation of research highlights from our preliminary study (results from Phase 1), the workshop was structured into two sessions: company presentations and group discussions. The workshop design was grounded in Kitzinger's principles on leveraging group dynamics in qualitative research \cite{kitzinger1995qualitative}, fostering dialogic exchange rather than isolated commentary.

\noindent\textbf{Company presentations} 
The workshop included eight presentations from six companies. Companies 2 and 4 gave two presentations, while the others presented once.
Each company's representatives were invited to respond to one or more guiding questions: (P1) What has your \emph{experience} been using generative AI in your test environments? (P2) What \emph{challenges} have you encountered in adopting generative AI techniques? and (P3) What are your \emph{hopes} for using generative AI to address these challenges?

\noindent\textbf{Cross-company group discussions} 
Participants were divided into five groups with cross-company representatives, each comprising 6–7 members. Each group was asked to answer the following questions: (G1) What are the priority research problems for industrial generative AI adoption? (G2) How should academia-industry collaborations be structured? (G3) What are the first steps for a small team of researchers working on this?
Finally, each group shared key discussion points with all attendees and submitted a summary to the authors. Note that the current paper provides an in-depth analysis only for the first question, while industry-academia collaboration aspects are considered out of scope.  

\subsection{Phase 3: Thematic coding analysis}
\label{sec:phase-3}

The data analyzed from the workshop consisted of two primary sources: (1) detailed notes taken during company presentations by all researchers, and (2) written summaries submitted by participants following group discussions. These materials from the workshop were analyzed using thematic coding analysis as described by Robson \cite{robson2024real}, with two researchers conducting the coding and the other two reviewing and refining the results, using the following five steps:

\begin{enumerate}
    \item \textbf{Familiarization with data:} The process began with repeatedly reading the collected notes to identify key patterns and noting initial ideas.
    
    \item \textbf{Generating initial codes:} Initial codes were developed by interacting inductively with the data to categorize recurring features. Similar extracts across the dataset were assigned the same codes.
    
    \item \textbf{Identifying and refining themes:} Codes were applied consistently to identical or closely related phenomena across different companies' presentations or group discussions. These codes were iteratively refined and collated into broader themes, with continuous comparison across the dataset.
    
    \item \textbf{Constructing thematic networks:} Emerging themes were grouped to form main or global themes.
    
    \item \textbf{Integration and interpretation:} The themes were compared with existing literature to identify gaps in academic knowledge regarding industrial needs.
\end{enumerate}

\section{Preliminary study} \label{sec:literature}

Based on the methodology for the preliminary study described in \autoref{sec:phase-1}, we reviewed research published after 2023 on how generative AI is utilized to test and simulate CPSs. To categorize and present this work, we used a hierarchical framework that captures levels of abstraction, tool categories, roles of large language models (LLMs), and associated techniques (see \autoref{tab:taxonomy}). Testing activities are often discussed according to test levels, including \emph{component}, \emph{integration}, and \emph{system} levels. Software tools used in engineering safety-critical CPS are categorized in tool qualification standards (like DO-330 \cite{DO330} used in civil avionics) as \emph{development} tools and \emph{verification} tools. In generative AI applications, LLMs can act in different roles depending on the task they are used for, such as \emph{reviewer}, \emph{documenter}, \emph{creator}, or \emph{translator}. Techniques used in this context include \emph{prompt engineering}, \emph{retrieval-augmented generation (RAG)}, \emph{LLM agents}, and \emph{fine-tuning}.

Considering existing approaches applied at the \emph{integration} and \emph{component} levels using \emph{development} tools, Chen et al. focused on deriving UML domain models from textual descriptions using different prompt strategies~\cite{DBLP:conf/models/ChenYCLMV23}. These approaches leveraged LLMs as \emph{creator}. Other studies emphasize the \emph{system}-level with  \emph{verification} tools. Xia et al. investigated the use of LLM-based multi-agent systems in digital twin environments to automatically adjust simulation parameters. Another example is Jackson et al., who worked on automating simulation scenarios and code generation from descriptions of systems and processes~\cite{jackson_natural_2024,xia_llm_2024}. Both studies leverage LLMs as \emph{translator} for high-level descriptions into executable simulations. Another example is Mühlburger and Wotawa, who utilized \emph{RAG} to detect faults in complex integrated systems~\cite{muhlburger_faultlines_2024}. Additionally, Deng et al. relied on natural language specifications of traffic rules and domain-specific languages for test scenarios and environments, along with prompting, to generate test scenarios and environment specifications~\cite{deng_target_2023}. Ali et al. studied the use of prompting to create digital twins for CPSs~\cite{ali_foundation_2025}. Across these efforts, LLMs are used mainly as \emph{creators} or \emph{reviewers}. Their results demonstrate how generative AI can help users better connect digital models with real-world systems.

\begin{table}[t]
\centering
\caption{A classification hierarchy illustrating examples from existing research.}
\label{tab:taxonomy}
\begin{adjustbox}{max width=\textwidth}
\rowcolors{2}{gray!15}{white}
\begin{tabular}{@{}c*{14}{P{0.7cm}}@{}}
\multicolumn{1}{c}{} &
\multicolumn{3}{c}{\textbf{Tool Category}} & 
\multicolumn{3}{c}{\textbf{Test Level}} & 
\multicolumn{4}{c}{\textbf{Role(s) of LLM}} & 
\multicolumn{4}{c}{\textbf{LLM Technique}} \\
\cmidrule(lr){2-4} \cmidrule(lr){5-7} \cmidrule(lr){8-11} \cmidrule(lr){12-15}
\rotatebox[origin=c]{270}{\textbf{Reference}} & 
\rotatebox[origin=c]{270}{} &
\rotatebox[origin=c]{270}{Development} & 
\rotatebox[origin=c]{270}{Verification} & 
\rotatebox[origin=c]{270}{Component} & 
\rotatebox[origin=c]{270}{Integration} & 
\rotatebox[origin=c]{270}{System} & 
\rotatebox[origin=c]{270}{Creator} & 
\rotatebox[origin=c]{270}{Translator} & 
\rotatebox[origin=c]{270}{Reviewer} & 
\rotatebox[origin=c]{270}{Documenter} & 
\rotatebox[origin=c]{270}{Prompting} & 
\rotatebox[origin=c]{270}{RAG} & 
\rotatebox[origin=c]{270}{LLM agents} & 
\rotatebox[origin=c]{270}{Fine-tuning} \\
\midrule
\cite{ali_foundation_2025}           &            &            & \checkmark &            &            & \checkmark & \checkmark &            &            & \checkmark & \checkmark &            &            &            \\
\cite{deng_target_2023}             &            &            & \checkmark &            &            & \checkmark &            &            & \checkmark &            & \checkmark &            &            &            \\
\cite{jackson_natural_2024}         &            &            & \checkmark &            &            & \checkmark &            & \checkmark &            &            &            &            & \checkmark &            \\
\cite{muhlburger_faultlines_2024}   &            &            & \checkmark &            &            & \checkmark &            &            & \checkmark &            &            & \checkmark &            & \checkmark \\
\cite{xia_llm_2024}                 &            &            & \checkmark &            &            & \checkmark &            & \checkmark &            &            &            &            & \checkmark &            \\
\cite{DBLP:conf/models/ChenYCLMV23} &            & \checkmark &            & \checkmark &            &            & \checkmark &            &            &            & \checkmark           &            &            &            \\
\bottomrule
\end{tabular}
\end{adjustbox}
\end{table}

In summary, recent studies predominantly focus on  \emph{verification} tools used at the \emph{system} level, with fewer addressing \emph{development} tools or targeting the \emph{component} and \emph{integration} levels. Most approaches rely on \emph{prompting} or are implemented via \emph{LLM agent} frameworks, while techniques like \emph{fine-tuning} are less frequently used. LLMs typically serve as \emph{creators} or \emph{translators}, with less emphasis on \emph{reviewer} or \emph{documenter} roles.

\section{Cross-company presentations and discussions} \label{sec:summaries}

This section summarizes the individual presentations delivered by each participating company, along with insights from cross-company group discussions held during the workshop. The cases span a range of industries, including automotive, communications, aerospace, and cybersecurity, and collectively, they offer a glimpse into how generative AI is transforming industrial test practices.

\subsection{Company presentations}

Below are summaries of the individual presentations delivered by company representatives during the workshop. These presentations were guided by the questions outlined in \autoref{sec:phase-2}. For anonymity, companies are labeled C1 to C6.

\vspace{2pt}
\noindent\textbf{C1}
Representatives from Company 1 reported their \emph{experience} in logging and interpreting real-time signals from truck components for autonomous vehicle development. The ten thousands of signals in modern vehicles require scalable data processing, which earlier state-machine-based approaches were unable to support. They also described efforts to automate build processes and testing to speed up feedback between customers and developers. Scaling to many signals presents \emph{challenges}, such as ensuring accuracy, organizing large log volumes, and adapting to unforeseen sensor needs. Transitioning to autonomous vehicles adds uncertainty in sensor placement and performance metrics due to varied driving patterns and specialized vehicle architectures. Looking ahead, Company 1 \emph{hopes} to leverage AI to utilize the vast amount of available data better, thereby shortening feedback loops and enhancing scalability in their products.

\vspace{2pt}
\noindent\textbf{C2} Engineers from Company 2 shared their \emph{experience} with simulation-based testing and large-scale data generation, highlighting its value as autonomous systems continue to evolve. Diverse teams use software-in-the-loop pipelines for validation and experiment with generative AI to build automatic driving controllers. Despite progress, Company 2 reports \emph{challenges} in maintaining model accuracy as simulations increase in scale and complexity. They face issues with reusing AI across environments, and language models show precision limits in complex simulations. Engineers express \emph{hope} for using AI to identify novel test scenarios, accelerate regression testing, and improve quality metrics. They see potential for AI to enhance realism in randomized simulations, support diverse scenario generation, and reduce the number of unproductive runs. Looking ahead, they aim to utilize AI to balance speed, quality, and resources in response to evolving industry demands.

\vspace{2pt}
\noindent\textbf{C3} A team from Company 3 did not report direct \emph{experience} with generative AI in simulators. However, other teams have developed simulation frameworks for emulating end users and network interactions. Notably, they shared experiences using generative AI for large-scale telecommunications testing, particularly in writing test cases, documentation, code refactoring, unit test generation, and vulnerability detection. They used retrieval-augmented generation to improve context awareness across codebases and align outputs with internal standards. They noted \emph{challenges} such as potential over-reliance on AI-generated code, which may lack quality or introduce vulnerabilities. Other concerns include intellectual property, security regulations, and difficulties integrating AI tools into existing CI pipelines. Standardizing prompt engineering across teams is also a barrier. They expressed \emph{hope} for more robust, developer-focused AI tools that complement human judgment. These tools could help meet data and security needs while improving efficiency in large-scale testing workflows.

\vspace{2pt}
\noindent\textbf{C4} Experts at Company 4 described their \emph{experience} in modeling and simulating complex subsystems using both physics-based and machine learning methods. They compared flight-derived data with simulated outputs to improve model fidelity and detect discrepancies between predictions and real-world behavior. They also used environmental simulation frameworks to advance sensor solutions, including neural model training for water dynamics and sea clutter. Simulated datasets feed into mathematical models, and internally developed architectures help distribute data to enhance simulations and create dynamic environments. Key \emph{challenges} include managing the vast volume of flight test data and converting it into coherent system-level metrics. Capturing subtle, high-level phenomena and modeling complex system-of-systems interactions remains difficult, especially as many simulator components are optimized for standalone rather than integrated performance. Still, they express \emph{hope} in AI-enabled “digital twin” techniques to align simulations with real flight behavior for more holistic system insights. They believe AI can enhance aircraft modeling, training realism, environmental understanding, and system interoperability.

\vspace{2pt}
\noindent\textbf{C5} Representatives from Company 5 did not report any \emph{experience} with generative AI in testing. They briefly described work on network segmentation and secure data diode installations. Their smaller size motivates more agile R\&D practices. The team identifies a \emph{challenge} in directing testing and quality assurance due to restricted data access and protected production environments. As their customers are security authorities, they must provide high levels of trust and proof of correctness, which increases the need for AI automation. These constraints necessitate innovative strategies that balance security with improved testing. They express a \emph{hope} that automated, accurate modeling of realistic customer environments using AI and digital twins will improve testing and build customer trust to share real-world data. They also see potential in AI-driven, scenario-based simulations for creating cost-effective and adaptable test environments. Combining customer data, simulation, and modeling is seen as invaluable.

\vspace{2pt}
\noindent\textbf{C6} Representatives from Company 6 have \emph{experience} in developing autonomous vehicle systems and applying AI-based testing. Their work included large-scale data logging from autonomous trucks and AI-based classification using a taxonomy that defines interests such as roundabouts, highways, or traffic density. This is a guided data-selection tool that sends chosen data for annotation to train AI models and create ground truth for validation. They also experimented with AI to generate challenging near-accident scenarios and build maps from logged data. Their main \emph{challenge} is managing massive data volumes and transitioning from rule-based simulations to dynamic generative AI environments. They face challenges in identifying critical data points, modeling near-accident scenarios, and scaling AI model training within a rigorous MLOps framework, necessitating ongoing refinement of simulation methods for safe and predictive autonomy. They express \emph{hope} in generative AI to create high-fidelity simulations. They anticipate that AI-generated maps, synthetic environments, and expanded MLOps capabilities will improve training and validation of autonomous systems.

\subsection{Cross-company group discussions} \label{sec:breakout}

This section summarizes the key points discussed by each group in response to the guiding question outlined in \autoref{sec:phase-2}, as collected according to the methodology described previously. The content is summarized but unfiltered, preserving the group-specific perspectives for later thematic coding analysis.

\vspace{2pt}
\noindent\textbf{Group 1} discussed the importance of applying generative AI to specific disciplines, noting that the value of results depends on context. They highlighted the challenge of detecting differences between simulations and reality, as well as handling hallucinations. The need for sensor input in test cases, especially in safety-critical testing, was also identified, as was the importance of using declassified, anonymized, and more representative synthetic data.

\vspace{2pt}
\noindent\textbf{Group 2} emphasized quality issues, including applicability, suitability, and scalability of generative AI solutions. They discussed both the solution domain (such as effective prompting, tracing, and distilling information) and the problem domain (development, testing, requirements, code, and test strategies). They questioned the extent to which these methods are applicable and suitable for their specific problems and applications. The group also explored the need for effective prompting techniques and noted that tracing information to artifacts is costly, suggesting that generative AI could help address this.

\vspace{2pt}
\noindent\textbf{Group 3} questioned whether AI-generated data and models can be trusted and how to prove their trustworthiness, especially when distributed AI systems are involved in handling process parts. They discussed creating machine learning models for agent (user) decision-making, such as those used in training or driving simulators. The group also mentioned the potential for generative AI to bridge the technical gap from requirements to code and its application in scheduling communication in simulators and degradation modeling. They reiterated the importance of ensuring that generated models are representative and relevant.

\vspace{2pt}
\noindent\textbf{Group 4} focused on modeling the surrounding environment for sensor simulations and using AI to select scenarios for best coverage or value, including edge cases. Given limited running resources, they discussed the need to evaluate scenario quality to choose among them. The group also raised the question of which metrics are most critical for evaluating or measuring simulation outcomes, and they mentioned the challenge of modeling environments with fidelity higher than current game engines.

\vspace{2pt}
\noindent\textbf{Group 5} addressed organizational set-up, specifically the need to bring together experts with domain knowledge and AI/ML experts. They raised the challenge of onboarding both groups effectively, noting that specialists understand the problem domain while technical experts may lack this context.

\section{Thematic coding analysis} \label{sec:thematic}

This section presents the results of the thematic coding analysis, organized into themes and sub-themes derived from recurring codes identified in the workshop data.
Such themes and sub-themes, with traceability to individual workshop presentations and group discussions, are presented in \autoref{tab:themes}. To contextualize the analysis, the table also includes a column referencing companies with reported experiences and related literature for each sub-theme. Note that our thematic coding analysis excludes aspects related to motivation and collaboration.

Next, each theme is discussed with direct references to corresponding statements or paraphrased excerpts that reflect the perspectives of industrial partners.

\subsection{AI-generated scenarios and environment models} \label{sec:ai_gen}

\begin{table}[tb]
\centering
\small
\renewcommand{\arraystretch}{1.3}
\caption{Summary of workshop themes and their mapping with literature}
\resizebox{\textwidth}{!}{%
\begin{tabular}{%
  >{\raggedright\arraybackslash}p{2.4cm}  
  >{\raggedright\arraybackslash}p{5.6cm} 
  >{\centering\arraybackslash}p{1.8cm}    
  >{\centering\arraybackslash}p{1.8cm}    
  >{\centering\arraybackslash}p{1.8cm}    
  >{\centering\arraybackslash}p{1.8cm}    
}
\toprule
\textbf{Theme} & \textbf{Sub-theme} & \multicolumn{2}{c}{\textbf{Hope/Challenge}} & \multicolumn{2}{c}{\textbf{Experience}} \\
 & & \textbf{Presentation} & \textbf{Group} & \textbf{Literature} & \textbf{Company} \\
\midrule

\multirow{3}{=}{AI-generated scenarios and environment models} 
  & \graycell Novel scenario generation       & \graycell C2, C3, C4, C5, C6 & \graycell G1, G4   & \graycell \cite{ali_foundation_2025}, \cite{deng_target_2023}, \cite{jackson_natural_2024}, \cite{muhlburger_faultlines_2024},  \cite{xia_llm_2024}    & \graycell C4, C6 \\
  & Environment model generation               & C4, C6             & G1, G3, G4             & \cite{ali_foundation_2025}, \cite{DBLP:conf/models/ChenYCLMV23}, \cite{deng_target_2023}, \cite{xia_llm_2024} & C4, C6 \\
  & \graycell Test scenario evaluation and selection & \graycell C6   & \graycell G4   & \graycell \cite{muhlburger_faultlines_2024}  & \graycell C6 \\
\hline
\multirow{3}{=}{Simulators and AI in CI/CD pipelines}
  & Large-scale data handling & C1, C4, C6   & -  & \cite{muhlburger_faultlines_2024}        & C1, C4, C6 \\
  & \graycell Continuous integration              & \graycell C1, C3, C6             & \graycell -             & \graycell -             & \graycell C1 \\
  & Simulators interoperability              & C2, C4             &  G3             &  -             &  C4 \\
\hline
\multirow{3}{=}{Trustworthiness in generative AI for simulation}
  & \graycell Model/Simulation fidelity                        & \graycell C2, C4             & \graycell G1, G3, G4             & \graycell \cite{ali_foundation_2025}, \cite{xia_llm_2024}                  & \graycell C4 \\
  & Over-reliance        & C3   & G1   & -        & - \\
  & \graycell Standardization and regulation      & \graycell C3, C5             & \graycell G2             & \graycell -             & \graycell C3 \\
  & Traceability & -   &  G2, G4   & -  & - \\

\bottomrule
\end{tabular}
}
\label{tab:themes}
\end{table}

Practitioners are exploring the use of generative AI to create scenarios and environment models. The following sub-themes have been identified:

\subsubsection{Novel scenario generation} involves using AI to generate new, diverse, or rare test scenarios that are challenging to capture with traditional, rule-based methods. The goal is to improve coverage in testing. Both automotive and aerospace companies emphasized the need to move beyond rigid frameworks and utilize AI to generate challenging or near-accident scenarios. For example, Company~2 aims to utilize AI to identify novel test scenarios and expedite regression testing, and Company~6 is experimenting with AI to create more challenging variations of near-accident scenarios. These efforts are echoed in group discussions, where the need for more representative and diverse synthetic data has been highlighted. The alignment across these reports shows a shared motivation to use generative AI for scenario creation.

\subsubsection{Environment model generation} involves using AI to build high-fidelity digital representations of real environments, such as flight conditions or customer networks. The aim is to bridge the gap between laboratory simulations and real-world testing. Companies~4 and~5 are working on creating robust virtual models, such as digital twins, to replicate real environments. Company~4 aims to utilize AI-enabled digital twin techniques to align real flight patterns with simulator outputs, while Company~5 recognizes the value in automated, accurate modeling of customer environments using AI and digital twins. Group~1 also mentioned the need for synthetic data generation. These examples demonstrate a consistent trend toward more realistic and adaptable simulation environments utilizing generative AI.

\subsubsection{Test scenario evaluation and selection} highlights the need for automated scenario evaluation methods and systems that can dynamically adapt the prompts fed into generative AI models, or even retrain the models themselves, in response to performance feedback. More importantly, we lack well-defined metrics to assess the quality of generated scenarios automatically. This is important for companies because they need it to select the most critical scenarios for running, given limited resources. For instance, Group~4 discussed the importance of evaluating scenario quality and choosing among them due to resource constraints. Company~6 uses AI-based classification to select test data and scenarios for annotation and model training. Current approaches reported in the literature~\cite{deng_target_2023,jackson_natural_2024,muhlburger_faultlines_2024,xia_llm_2024} often rely on manual validation of generated scenarios, which is not scalable given the industry's need to use large, complex datasets.

\subsection{Simulators and AI in CI/CD pipelines} \label{sec:sim_ci}

This theme addresses the technical challenges of integrating AI-based approaches into existing development and testing pipelines, especially at scale.

\subsubsection{Large-scale data handling} refers to the need for scalable data processing and management as data volumes and complexity grow, particularly in autonomous systems. Companies~1, 4, and~6 all report handling large amounts of data from real-world operations, such as autonomous trucks or flight tests. Company~1 highlights the need for scalable data processing to handle tens of thousands of signals, while Company~4 focuses on converting vast test data into actionable metrics. Company~6 logs large-scale data from autonomous vehicles. These experiences highlight a common challenge in managing and analyzing complex, high-volume data streams, underlining the need for robust data pipelines.

\subsubsection{Continuous integration (CI)} focuses on integrating AI-driven solutions into CI and automated workflows, including regular model retraining and dataset updates. Companies~1 and~3, and Company~6, all emphasize the importance of integrating AI into CI pipelines. Company~1 invests in automated build and testing cycles to speed up feedback, while Company~3 faces challenges adapting AI tools to established CI processes. Company~6 is scaling AI model training within a rigorous MLOps framework. These reports highlight the need for seamless integration of AI into existing development cycles.

\subsubsection{Simulators interoperability} covers the challenge of integrating multiple simulators or subsystems, often designed to operate independently, into a cohesive system for end-to-end validation. Company 4 reports difficulties in connecting separate simulation frameworks for system-of-systems validation, noting that complex interactions between subsystems are challenging to model. Their internal solutions aim to make data accessible across the organization. This aligns with the need for better communication and data sharing between simulators, as discussed in group sessions.

\subsection{Trustworthiness in generative AI for simulation} \label{sec:trust}

This theme focuses on ensuring that AI-generated data, models, and simulations are reliable, accurate, and compliant with standards and regulations.

\subsubsection{Model/Simulation fidelity} is about ensuring that simulations and models accurately reflect real-world behavior, especially in complex or safety-critical systems. Companies~2,~4, and~6, along with several groups, highlighted the challenge of maintaining model fidelity as complexity increases. Company~2 faces difficulties maintaining accuracy as simulation complexity grows, while Company~4 reported their experience in comparing real flight data with simulations to set credibility benchmarks. Group 1 discussed the challenge of distinguishing between simulated and real-world scenarios. These aligned reports show that simulation fidelity is a concern across domains.

\subsubsection{Over-reliance} refers to the risk of depending too much on AI-generated outputs, which may be of uncertain quality or introduce errors. Company~3 warns of potential over-reliance on AI-suggested code, while Group~1 raises concerns about handling hallucinations in AI-generated test cases, especially in safety-critical contexts. Both point to the need for careful oversight.

\subsubsection{Standardization and regulation} addresses the need for internal standards and regulatory compliance when using AI-generated artifacts, particularly in terms of intellectual property and data security. Companies~3 and~5 express concerns about intellectual property, data security, and regulatory restrictions when using AI-generated code or test artifacts. Company~3 also notes the challenge of standardizing prompt engineering across teams. Group~2 discussed the need for effective prompting. These reports show a shared need for clear standards and regulatory guidance.

\subsubsection{Traceability} aims to link information to artifacts or find relevant requirements. Group~2 suggested that generative AI could assist with such costly tasks.

\subsection{Summary and analysis}

The thematic coding analysis presented here in \autoref{sec:thematic} resulted in three major themes, representing the three areas where generative AI techniques have the most potential for simulations in test environments for large-scale cyber-physical systems, according to the practitioners at the cross-company workshop. 
Those three areas are AI-generated scenarios and environment models, Simulators and AI in CI/CD pipelines, and Trustworthiness in generative AI for simulation.

Participants from several companies at the cross-company workshop described the need to increase efficiency and effectiveness in their test activities, exemplified by reducing manual work, improving test coverage, shortening feedback loops, and implementing automated testing. For example, one participant described how the company aims to “exploit techniques to harness the vast amount of data already generated”, potentially shortening the feedback loop. A participant from another company described how their company sees advanced AI techniques as a means to enhance realism and maintain simulation ecosystems in alignment with rapidly evolving industry demands. In analyzing the results from the cross-company workshop, we found that companies see potential in generative AI for simulation, but do not appear to have a clear agenda going forward, suggesting a need for continued industry-academia collaboration.

\section{Threats to validity} \label{sec:threats}

\noindent\textbf{Construct validity} One must always consider that a different set of questions and context for the cross-company workshop can lead to a different focus for the participants. To address threats to construct validity, the questions for the presentations and breakout sessions were designed with open-ended questions. In this paper, we also present background material for both workshop participants and the companies in the study, providing as much information as possible about the context to enable the reproducibility of the study.

Another threat to construct validity is researcher bias while interpreting the results from the cross-company workshop. To mitigate this threat, two researchers analyzed the data separately, and the other two reviewed the results to ensure quality and correctness. The process was conducted iteratively to enhance the quality of the analysis, achieving consensus through discussions and visualizations in diagrams, tables, and text.

\noindent\textbf{External validity} The cross-company workshop included participants from six companies. Due to this, it is conceivable that the ﬁndings from this study are only valid for these companies or companies operating in the same industry segments (presented in \autoref{sec:phase-2}). The diverse nature of these six companies ensures that our study covers a variety of industrial CPSs. As such, it is reasonable to expect that the study's results are also relevant to a large segment of the CPS industry (analytic generalization). However, we consider external validation in other companies (preferably in different industry segments) as future work.

\noindent\textbf{Internal validity} Of the 12 threats to internal validity listed by Cook, Campbell, and Day \cite{cook1979quasi}, we consider selection, ambiguity about causal direction, and compensatory rivalry relevant to this work:

\begin{itemize}
    \item \emph{Selection}: All workshop participants were purposively sampled (selected as good informants with appropriate roles in the companies) according to the guidelines for qualitative data appropriateness provided by Robson \cite{robson2024real}. Based on the rationale of these samplings and supported by Robson, who considers this type of sampling superior for this type of study to ensure appropriateness, we consider this threat to be mitigated.
    \item \emph{Ambiguity about causal direction}: While this study discusses relationships in some cases, we are cautious about making statements regarding causation. Statements that include cause and effect are collected from the presentations and break-out sessions at the cross-company workshop and not introduced in the interpretation of the data.
    \item \emph{Compensatory rivalry}: In qualitative research, the threat of compensatory rivalry must always be considered. The questions for the presentations and breakout sessions were deliberately designed to be value-neutral for the participants, avoiding judgment of their performance or skills. Generally, the questions were also designed to be open-ended, preventing bias and ensuring open and accurate answers. However, our experiences from previous work indicate that participants are more prone to self-criticism than to self-praise.
\end{itemize}

\section{Conclusions and future work} \label{sec:conclusion}

Our cross-company study reveals that while the industry sees potential in generative AI for more scalable and efficient simulation and testing, several key areas require further research for its effective adoption in simulation-based testing. According to practitioners, the three areas where generative AI techniques have the most potential for simulations in test environments for large-scale CPSs are:

\begin{itemize}
    \item \textbf{AI-generated scenarios and environment models:} This line of work applies generative AI to create and enhance simulation scenarios and environment models, and to define clear metrics to assess these artifacts by novelty and realism. Achieving this demands strong domain-specific insight and a deep understanding of the operational context.
    
    \item \textbf{Simulators and AI in CI/CD pipelines:} This research direction investigates how generative AI and simulation tools can be integrated into CI/CD pipelines to enable scalable data processing, automated model retraining, and orchestrated validation across multi-disciplinary systems or simulators.
    
    \item \textbf{Trustworthiness in generative AI for simulation:} Research in this direction is about ensuring that AI-generated simulations and models are trustworthy enough for safety-critical systems or production use. It verifies that simulations accurately reflect real-world behavior, establishes systematic oversight and traceability, and aligns outputs with internal standards and external regulations.
\end{itemize}

While companies seem to lack a clear, actionable agenda for realizing the potential of generative AI in the context of CPSs, our identified areas faithfully capture various opportunities and challenges that participating companies face. 

These areas can also serve as a starting point for strengthened industry-academia collaboration, primarily as companies strive to transform large-scale, complex operational data into actionable insights, as mentioned by multiple companies. Participants repeatedly emphasized the value of such collaborations and the importance of tangible outputs to sustain engagement, highlighting the need to deliver visible results and proofs of concept to demonstrate concrete feasibility and value and secure managerial buy-in.

Overall, current generative AI solutions appear more suited for one-off or exploratory use cases rather than integration into production environments, where higher levels of automation and trustworthiness are essential. This was reflected in two key concerns raised by practitioners: the lack of reliable evaluation mechanisms to ensure trustworthy outputs, and the difficulty of integrating generative components into continuous integration and deployment pipelines. When revisiting the literature from the preliminary study phase, we found that these concerns were largely unaddressed, and many approaches relied on manual evaluation and expert-driven prompt or parameter tuning, making them inadequate for scalable, automated deployment. These persistent limitations in both practice and research highlight the need for further development of robust evaluation methods and seamless pipeline integration. Moreover, the anticipated applications of generative AI outlined here remain future work, and structured discussions with experts in the field of generative AI will be essential to assess their feasibility.

\subsubsection*{Acknowledgment}
We sincerely appreciate Software Center for organizing the workshop and the participating companies for generously sharing their experiences and insights. The first author was partially supported by the Vinnova competence center on Continuous Digitalization (CoDiG), while the last author was partially supported by the Wallenberg AI, Autonomous Systems and Software Program (WASP) funded by the Knut and Alice Wallenberg Foundation. The authors grateful to Willem Meijer for his feedback on the manuscript draft.

\bibliographystyle{splncs04}
\bibliography{bibs}

@inproceedings{DBLP:conf/models/ChenYCLMV23,
  author       = {Kua Chen and
                  Yujing Yang and
                  Boqi Chen and
                  Jos{\'{e}} Antonio Hern{\'{a}}ndez L{\'{o}}pez and
                  Gunter Mussbacher and
                  D{\'{a}}niel Varr{\'{o}}},
  title        = {Automated Domain Modeling with Large Language Models: {A} Comparative
                  Study},
  OPTbooktitle    = {26th {ACM/IEEE} International Conference on Model Driven Engineering
                  Languages and Systems, {MODELS} 2023, V{\"{a}}ster{\aa}s, Sweden,
                  October 1-6, 2023},
  booktitle    = {26th {ACM/IEEE} Int. Conf. on Model Driven Engineering
                  Languages and Systems, {MODELS} 2023, V{\"{a}}ster{\aa}s, Sweden,
                  October 1-6, 2023},
  pages        = {162--172},
  publisher    = {{IEEE}},
  year         = {2023},
  doi          = {10.1109/MODELS58315.2023.00037},
  bibsource    = {dblp computer science bibliography, https://dblp.org}
}

@inproceedings{DBLP:conf/vecos/PatilUN24,
  author       = {Minal Suresh Patil and
                  Gustav Ung and
                  Mattias Nyberg},
  OPTeditor       = {Bernhard Steffen},
  title        = {Towards Specification-Driven LLM-Based Generation of Embedded Automotive
                  Software},
  booktitle    = {Bridging the Gap Between {AI} and Reality - Second Int. Conf.,
                  AISoLA 2024, Crete, Greece, October 30 - November 3, 2024},
  series       = {Lecture Notes in Computer Science},
  volume       = {15217},
  pages        = {125--144},
  publisher    = {Springer},
  year         = {2024},
  doi          = {10.1007/978-3-031-75434-0\_9},
  bibsource    = {dblp computer science bibliography, https://dblp.org}
}

@inproceedings{ali_foundation_2025,
  author       = {Shaukat Ali and
                  Paolo Arcaini and
                  Aitor Arrieta},
  OPTeditor       = {Tiziana Margaria and
                  Bernhard Steffen},
  title        = {Foundation Models for the Digital Twins Creation of Cyber-Physical
                  Systems},
  OPTbooktitle    = {Leveraging Applications of Formal Methods, Verification and Validation.
                  Application Areas - 12th International Symposium, ISoLA 2024, Crete,
                  Greece, October 27-31, 2024, Proceedings, Part {V}},
  booktitle    = {Leveraging Applications of Formal Methods, Verification and Validation.
                  Application Areas - 12th Int. Symposium, ISoLA 2024 Part {V}},
  OPTseries       = {Lecture Notes in Computer Science},
  series       = {LNCS},
  volume       = {15223},
  pages        = {9--26},
  publisher    = {Springer},
  year         = {2024},
  doi          = {10.1007/978-3-031-75390-9\_2},
  bibsource    = {dblp computer science bibliography, https://dblp.org}
}

@book{cook1979quasi,
  title={Quasi-experimentation: Design \& analysis issues for field settings},
  author={Cook, Thomas D and Campbell, Donald Thomas and Day, Arles},
  volume={351},
  year={1979},
  publisher={Houghton Mifflin Boston}
}

@misc{deng_target_2023,
	title = {{TARGET}: Automated scenario generation from traffic rules for testing autonomous vehicles},
	doi = {10.48550/arXiv.2305.06018},
	shorttitle = {generate test scenarios for ads from traffic rules},
	author = {Deng, Yao and Yao, Jiaohong and Tu, Zhi and Zheng, Xi and Zhang, Mengshi and Zhang, Tianyi},
	urldate = {2024-10-10},
	date = {2023-05-01},
	langid = {american},
}

@article{jackson_natural_2024,
  author       = {Ilya Jackson and
                  Mar{\'{\i}}a Jes{\'{u}}s S{\'{a}}enz and
                  Dmitry A. Ivanov},
  title        = {From natural language to simulations: applying {AI} to automate simulation
                  modelling of logistics systems},
  journal      = {Int. J. Prod. Res.},
  volume       = {62},
  number       = {4},
  pages        = {1434--1457},
  year         = {2024},
  doi          = {10.1080/00207543.2023.2276811},
  bibsource    = {dblp computer science bibliography, https://dblp.org}
}

@article{kitzinger1995qualitative,
  title={Qualitative research: introducing focus groups},
  author={Kitzinger, Jenny},
  journal={Bmj},
  volume={311},
  number={7000},
  pages={299--302},
  year={1995},
  publisher={British Medical Journal Publishing Group}
}

@inproceedings{maartensson2017continuous,
  author       = {Torvald M{\aa}rtensson and
                  Daniel St{\aa}hl and
                  Jan Bosch},
  title        = {Continuous integration impediments in large-scale industry projects},
  booktitle    = {2017 {IEEE} Int. Conf. on Software Architecture, {ICSA}
                  2017, Gothenburg, Sweden, April 3-7, 2017},
  pages        = {169--178},
  publisher    = {{IEEE} Computer Society},
  year         = {2017},
  doi          = {10.1109/ICSA.2017.11},
  bibsource    = {dblp computer science bibliography, https://dblp.org}
}

@article{https://doi.org/10.1002/stvr.1839,
  author       = {Torvald M{\aa}rtensson and
                  G{\"{o}}ran Ancher and
                  Daniel St{\aa}hl},
  title        = {Test environments for large-scale software systems - An industrial
                  study of intrinsic and extrinsic success factors},
  journal      = {Softw. Test. Verification Reliab.},
  volume       = {33},
  number       = {3},
  year         = {2023},
  doi          = {10.1002/STVR.1839},
  bibsource    = {dblp computer science bibliography, https://dblp.org}
}

@inproceedings{muhlburger_faultlines_2024,
  author       = {Herbert M{\"{u}}hlburger and
                  Franz Wotawa},
  title        = {FaultLines - Evaluating the Efficacy of Open-Source Large Language
                  Models for Fault Detection in Cyber-Physical Systems},
  OPTbooktitle    = {{IEEE} International Conference on Artificial Intelligence Testing,
                  AITest 2024, Shanghai, China, July 15-18, 2024},
  booktitle    = {{IEEE} Int. Conf. on Artificial Intelligence Testing,
                  AITest 2024},
  pages        = {47--54},
  publisher    = {{IEEE}},
  year         = {2024},
  doi          = {10.1109/AITEST62860.2024.00014},
  bibsource    = {dblp computer science bibliography, https://dblp.org}
}

@book{robson2024real,
  title={Real world research},
  author={Robson, Colin},
  year={2024},
  publisher={John Wiley \& Sons},
  edition={5}
}

@inproceedings{wohlin2014guidelines,
  author       = {Claes Wohlin},
  OPTeditor       = {Martin J. Shepperd and
                  Tracy Hall and
                  Ingunn Myrtveit},
  title        = {Guidelines for snowballing in systematic literature studies and a
                  replication in software engineering},
  booktitle    = {18th Int. Conf. on Evaluation and Assessment in Software
                  Engineering, {EASE} '14, London, England, United Kingdom, May 13-14,
                  2014},
  pages        = {38:1--38:10},
  publisher    = {{ACM}},
  year         = {2014},
  doi          = {10.1145/2601248.2601268},
  bibsource    = {dblp computer science bibliography, https://dblp.org}
}

@inproceedings{xia_llm_2024,
  author       = {Yuchen Xia and
                  Daniel Dittler and
                  Nasser Jazdi and
                  Haonan Chen and
                  Michael Weyrich},
  title        = {{LLM} experiments with simulation: Large language model multi-agent
                  system for simulation model parametrization in digital twins},
  OPTbooktitle    = {29th {IEEE} International Conference on Emerging Technologies and
                  Factory Automation, {ETFA} 2024, Padova, Italy, September 10-13, 2024},
  booktitle    = {29th {IEEE} Int. Conf. on Emerging Technologies and
                  Factory Automation, {ETFA} 2024},
  pages        = {1--4},
  publisher    = {{IEEE}},
  year         = {2024},
  doi          = {10.1109/ETFA61755.2024.10710900},
  bibsource    = {dblp computer science bibliography, https://dblp.org}
}

@article{DBLP:journals/tcps/LeeCJKHUKP25,
  author       = {Sanghoon Lee and
                  Jiyeong Chae and
                  Haewon Jeon and
                  Taehyun Kim and
                  Yeong{-}Gi Hong and
                  Doo{-}Sik Um and
                  Taewoo Kim and
                  Kyung{-}Joon Park},
  title        = {Cyber-Physical {AI:} Systematic Research Domain for Integrating {AI}
                  and Cyber-Physical Systems},
  journal      = {{ACM} Trans. Cyber Phys. Syst.},
  volume       = {9},
  number       = {2},
  pages        = {19:1--19:33},
  year         = {2025},
  doi          = {10.1145/3721437},
  bibsource    = {dblp computer science bibliography, https://dblp.org}
}

@article{DBLP:journals/spe/CederbladhEMS24,
  author       = {Johan Cederbladh and
                  Romina Eramo and
                  Vittoriano Muttillo and
                  Per Erik Strandberg},
  title        = {Experiences and challenges from developing cyber-physical systems
                  in industry-academia collaboration},
  journal      = {Softw. Pract. Exp.},
  volume       = {54},
  number       = {6},
  pages        = {1193--1212},
  year         = {2024},
  doi          = {10.1002/SPE.3312},
  bibsource    = {dblp computer science bibliography, https://dblp.org}
}

@article{DBLP:journals/corr/abs-2501-04410,
  author       = {Krisztian Balog and
                  ChengXiang Zhai},
  title        = {User Simulation in the Era of Generative {AI:} User Modeling, Synthetic
                  Data Generation, and System Evaluation},
  journal      = {CoRR},
  volume       = {abs/2501.04410},
  year         = {2025},
  doi          = {10.48550/ARXIV.2501.04410},
  eprinttype    = {arXiv},
  eprint       = {2501.04410},
  bibsource    = {dblp computer science bibliography, https://dblp.org}
}

@article{DBLP:journals/corr/abs-2406-17112,
  author       = {Kassi Muhammad and
                  Teef David and
                  Giulia Nassisid and
                  Tina Farus},
  title        = {Integrating Generative {AI} with Network Digital Twins for Enhanced
                  Network Operations},
  journal      = {CoRR},
  volume       = {abs/2406.17112},
  year         = {2024},
  doi          = {10.48550/ARXIV.2406.17112},
  eprinttype    = {arXiv},
  eprint       = {2406.17112},
  bibsource    = {dblp computer science bibliography, https://dblp.org}
}

@article{DBLP:journals/access/ZhouGHY18,
  author       = {Xin Zhou and
                  Xiaodong Gou and
                  Tingting Huang and
                  Shunkun Yang},
  title        = {Review on testing of cyber physical systems: methods and testbeds},
  journal      = {{IEEE} Access},
  volume       = {6},
  pages        = {52179--52194},
  year         = {2018},
  doi          = {10.1109/ACCESS.2018.2869834},
  bibsource    = {dblp computer science bibliography, https://dblp.org}
}

@inproceedings{DBLP:conf/profes/KarlssonLSS24,
  author       = {Albin Karlsson and
                  Erik Lindmaa and
                  Simin Sun and
                  Miroslaw Staron},
  OPTeditor       = {Dietmar Pfahl and
                  Javier Gonzalez{-}Huerta and
                  Jil Kl{\"{u}}nder and
                  Hina Anwar},
  title        = {{AI}-Based Automotive Test Case Generation: An Action Research Study
                  on Integration of Generative {AI} into Test Automation Frameworks},
  booktitle    = {Product-Focused Software Process Improvement. Industry-, Workshop-,
                  and Doctoral Symposium Papers - 25th Int. Conf., {PROFES}
                  2024, Tartu, Estonia, December 2-4, 2024},
  series       = {LNCS},
  volume       = {15453},
  pages        = {50--66},
  publisher    = {Springer},
  year         = {2024},
  doi          = {10.1007/978-3-031-78392-0\_4},
  bibsource    = {dblp computer science bibliography, https://dblp.org}
}

@article{zhuo2024bigcodebench,
  title={Bigcodebench: Benchmarking code generation with diverse function calls and complex instructions},
  author={Zhuo, Terry Yue and Vu, Minh Chien and Chim, Jenny and Hu, Han and Yu, Wenhao and Widyasari, Ratnadira and Yusuf, Imam Nur Bani and Zhan, Haolan and He, Junda and Paul, Indraneil and others},
  journal={arXiv preprint arXiv:2406.15877},
  year={2024}
}

@inproceedings{alshahwan2024automated,
  title={Automated unit test improvement using large language models at meta},
  author={Alshahwan, Nadia and Chheda, Jubin and Finogenova, Anastasia and Gokkaya, Beliz and Harman, Mark and Harper, Inna and Marginean, Alexandru and Sengupta, Shubho and Wang, Eddy},
  booktitle={Companion Proceedings of the 32nd ACM Int. Conf. on the Foundations of Software Engineering},
  OPTbooktitle={Companion Proceedings of the 32nd ACM International Conference on the Foundations of Software Engineering},
  pages={185--196},
  year={2024}
}

@misc{DO330,
  author       = {{RTCA}},
  title        = {{DO-330 Software Tool Qualification Considerations}},
  year         = {2011},
  organization = {RTCA, Inc.},
  note         = {Available from: \url{https://my.rtca.org/productdetails?id=a1B36000001IcfkEAC}},
}

\end{document}